\documentclass{article}
\pdfoutput=1
\usepackage{natbib}
\usepackage{amsmath}
\usepackage{amssymb}
\usepackage{booktabs}
\usepackage{dsfont}
\usepackage{slashed}
\usepackage{multicol}
\usepackage{caption}
\usepackage[top=3cm,bottom=3.5	cm,left=2.75cm,right=2.75cm]{geometry}
\usepackage{graphicx}

\newenvironment{Figure}
  {\par\medskip\noindent\minipage{\linewidth}}
  {\endminipage\par\medskip}

\usepackage{abstract}

\usepackage{titlesec}
\titleformat{\section}[block]{\large\scshape\centering}{\thesection.}{1em}{}
\titleformat{\subsection}[block]{\scshape}{\thesubsection.}{1em}{}
\title{
	\fontsize{20pt}{5pt}\selectfont
	\textbf{GAz: A Genetic Algorithm for Photometric Redshift Estimation}
	}	
\author{
	\large
	\textsc{Robert Hogan\footnote{robert.hogan@kcl.ac.uk}, Malcolm Fairbairn\footnote{malcolm.fairbairn@kcl.ac.uk}, and Navin Seeburn \footnote{navin.seeburn.2012@live.rhul.ac.uk. Current address: Royal Holloway, University of London, Egham, TW20 0EX, UK}}
	\\[2mm]
	\normalsize	Physics, Kings College London, Strand, London WC2R 2LS, UK
	\vspace{-.5cm}
	}
\date{}

\begin{document}
\maketitle
\begin{abstract}
\noindent We present a new approach to the problem of estimating the redshift of galaxies from photometric data. The approach uses a genetic algorithm combined with non-linear regression to model the 2SLAQ LRG data set with SDSS DR7 photometry. The genetic algorithm explores the very large space of high order polynomials while only requiring optimisation of a small number of terms. We find a $\sigma_{\text{rms}}=0.0408\pm 0.0006$ for redshifts in the range $0.4<z< 0.7$. These results are competitive with the current state-of-the-art but can be presented simply as a  polynomial which does not require the user to run any code. We demonstrate that the method generalises well to other data sets and redshift ranges by testing it on SDSS DR11 and on simulated data. For other datasets or applications the code has been made available at https://github.com/rbrthogan/GAz.
\end{abstract}
\vspace{.5cm}
\begin{multicols}{2}
\section{Introduction}
Large scale structure cosmology and extragalactic astronomy rely heavily on accurate estimate of the redshift of objects under study. For example the reconstruction of the two point correlation function for galaxies is critical to understand the history of structure formation in the Universe and probe theories beyond $\Lambda$CDM. Unfortunately it is a very time consuming and expensive task to obtain spectroscopic data for the millions of observed galaxies. It has therefore long been a challenge to estimate the redshift of galaxies using the much easier to obtain photometric data.
\subsection{Photometric Redshifts}
The estimation of redshifts from photometric data has been an industry for some years in astronomy culminating in the production of MegaZ-LRG \citep{Collister07}, a database of photometric redshifts of 1 million luminous red galaxies in the range $0.4<z<0.7$ and the 2MPZ database \citep{2MPZ1,2MPZ2} for $z<0.3$. The aim is to model the spectroscopic redshift using photometric redshift estimator, $z_{\text{phot}}(u,g,r,i,z)$, where $\{u, g,r,i,z\}$ are the standard photometric magnitudes. There have been two main approaches to the problem: template based methods \citep{BPz,EAZY,GAZELLE,GOODZ,Hyperz, LePhare1,LePhare2,LRT1,LRT2,ZEBRA} and machine learning/empirical methods \citep{ANNz,ArborZ,Wolf09,Csabai2007,RandomForests,Brescia2014,GLM}. For comparisons of the various codes see \citep{Abdalla0812,Hildebrandt1008,Dahlen1308}. One of the best performing codes is ANN$z$\citep{ANNz} which is based on artificial neural networks and was used in creating the MegaZ and 2MPZ databases. 

With so many well preforming codes already available one may ask if it is useful to introduce yet another solution. Many applications of photometric redshift catalogues are, however, still limited by the imperfect reconstruction see (e.g. \cite{Asorey1305} for an application of photometric redshifts to measurements of redshift-space distortions that is limited by photometric redshift error). There is therefore still a need to further reduce the error in photometric redshifts and that is the aim of this paper.

\subsection{Genetic Algorithms}
Genetic algorithms (GAs) \citep{holland75,goldberg89,reeves03} are a non-derivative optimisation method inspired by evolution by natural selection. They can be very useful for optimisation problems on sets and spaces that are discrete or very large. The former two examples can be very difficult to tackle with traditional optimisation algorithms that rely on derivatives or require a distance measure on the space. If, for example, a space doesn't have a distance measure then algorithms that require 'small steps' are meaningless. GAs, however, represent points in these spaces by a genetic encoding and explore the space by evolving a population of these genomes together. The algorithm proceeds as you would expect from evolutionary biology:
\begin{enumerate}
\item select genetic encoding for trial solutions (e.g. binary),
\item generate initial random population of solutions using encoding,
\item evaluate performance of each individual using some fitness function,
\item select individuals for breeding with preferential selection for fitter individuals,
\item breed selected individuals by combining genetic information in some way to create offspring,
\item introduce random mutation in offspring to produce new generation,
\item repeat steps 3-6 to evolve population towards optimal solution.
\end{enumerate}
As a simple example, the canonical GA (which uses a binary encoding) could be used to maximize a function $f(x)$. A trial solution would be a binary encoding of a floating point number (the number of bits would be determined by the precision required for the solution). The fitness would then be determined by evaluating $f(x)$ for each individual (larger values are deemed to be fitter). We could then select the fittest half of the population and (using two copies of each) breed these solutions by splitting the bit strings a random point and swapping the ends to produce two offspring. A mutation would be a simple bit flip of the individual. As the algorithm iterates through many generations the solutions would approach the maximum. This is of course not a very good use of the power of a GA because we have for more appropriate methods to maximise continuous functions, especially if we are able to compute a derivative. There are however other evolutionary algorithms (e.g. differential evolution \citep{Storn97}) that are well suited to this task.

GAs have found application in areas such as bioinformatics, schedule optimisation, automated design, and game theory. In a physics context, examples of recent use of evolutionary algorithms include SUSY model selection \citep{Allanach0406}, string landscape exploration \citep{Abel1404}, and event selection for particle physics experiments \citep{Cranmer0402}.

In this paper we use a GA to determine the optimal polynomial form to fit to redshift training data. We use this method to reduce the photometric redshift error on the 2SLAQ data set. The rest of the paper is organised as follows: in section 2 we describe the algorithm and optimise it for our purposes, in section 3 we present our results, and in section 4 we conclude.
\section{The Algorithm}
\subsection{Overview}
In our case, we want to model a function of 5 input variables $z_{\text{phot}}(u,g,r,i,z)$. We restrict the form of $z_{\text{phot}}$ to be a polynomial degree, $N$, and limit the number of terms, $t$,
\begin{equation}
z_{\text{phot}}=\sum^t_{j=1}a_j u^{k_j} g^{l_j}  r^{m_j}  i^{n_j}  z^{o_j} ,
\end{equation}
with $k_j+l_j+m_j+n_j+o_j \leq N$. For fixed $N$ and $t$, the task of the GA is then to find the optimum choice of polynomial to produce an accurate model for $z_{\text{phot}}$. Each individual therefore represents a particular polynomial, and the fitness of that individual is then determined by fitting that polynomial (i.e. the $a_j$'s) to the data. Our genetic encoding is therefore \footnote{We follow the encoding of \citep{Clegg05} that first used this approach for noiseless data fitting.}
\begin{align}
[[ k_1, l_1, m_1,n_1,o_1] , [ k_2 &, l_2, m_2,n_2,o_2 ],\label{eqn: genetic encoding}\\
&...,[ k_t, l_t, m_t,n_t,o_t]]. \nonumber
\end{align}
As an example, if we have $N=3$ and $t=3$ an possible individual could be 
\begin{equation}
[[0,0,0,0,0],[1,0,0,0,0],[0,2,1,0,0]], 
\end{equation}
which translates to the polynomial
\begin{equation}
a_1+ a_2 u +a_3 g^2 r.
\end{equation}
The coefficients $a_1,\ a_2,\ a_3$ would then be determined by fitting to the training data.

We generate an initial population of $P$ individuals ensuring that there are no repeated terms in any individual. The breeding process takes two individuals of the form of Eqn. (\ref{eqn: genetic encoding}) and randomly shuffles the terms to create two offspring (again ensuring no repeat terms). In order to mutate an individual each term is replaced by a new term with some small probability, $p_m$. An alternative approach to mutation would be to mutate on the subterm level i.e. mutate the $\{k_j,l_j,m_j,n_j,o_j \}$ individually. This approach will likely result in a slower, more controlled search, but given the combinatorially large search space the success of this approach would likely depend too strongly on the original population. 

We are trying to optimise the fit of the spectroscopic redshift, $z_{\text{spec}}$, so we minimize the least-squares error,
\begin{equation}
E=\sum_j^n (z_{\text{spec}}^j-z_{\text{phot}}^j(u,g,r,i,z))^2,
\end{equation}
where $n$ is the total number of galaxies in the training set and we define our fitness as $f=1/E$. The are several possible choices for $f$ and depending on the selection method used they can have an effect on performance the GA. In the next section we optimise the GA for this particular choice of fitness function. There is a notable absence of the experimental errors in the this fitness function. This is unfortunate but necessary and we will return to this point later.

\subsection{GA Optimisation}
In order to get the best performance from a GA it is important to tune the various hyper-parameters (selection mechanism, mutation rate, etc.) for your particular application. There are some general guidelines for making these choices in the case of the canonical GA but the exercise must be repeated for different genetic encodings as is the case in this paper. 
\subsubsection*{Selection Mechanism}
Perhaps the most significant difference between different GAs is in the choice of selection method. The optimal choice for each application will depend largely on how you implement your fitness function. Selection methods can be broadly classified into the fitness proportionate and rank based selection. Different methods have different levels of selection pressure i.e. some methods allow only the very fittest to breed where as others give weaker individuals more of a chance in order to maintain genetic diversity in the population.

Fitness proportionate selection simply weight the probably of an individual being selected for breeding by its fitness relative to the average fitness. Individuals with a very large probability are likely to be selected several times. An efficient implementation of fitness proportionate selection is what is called roulette wheel selection that arranges individuals into a roulette wheel with area of each segment weighted by the fitness of the individual. The outside of the wheel has $P$ equally spaced marks. A single random number generation (or spin of the wheel) aligns the marks with segments and simultaneously selects $P$ breeding candidates proportional to the fitness of the population.

Another selection method is tournament selection. In this case, $P$ random fixed size subsamples of the population are drawn and the fittest candidate from each subsample is selected for breeding. The size of the tournaments can be varied to tune the selection pressure with larger tournaments penalising weaker individuals more.
\begin{Figure}
\centering
\includegraphics[width=\linewidth]{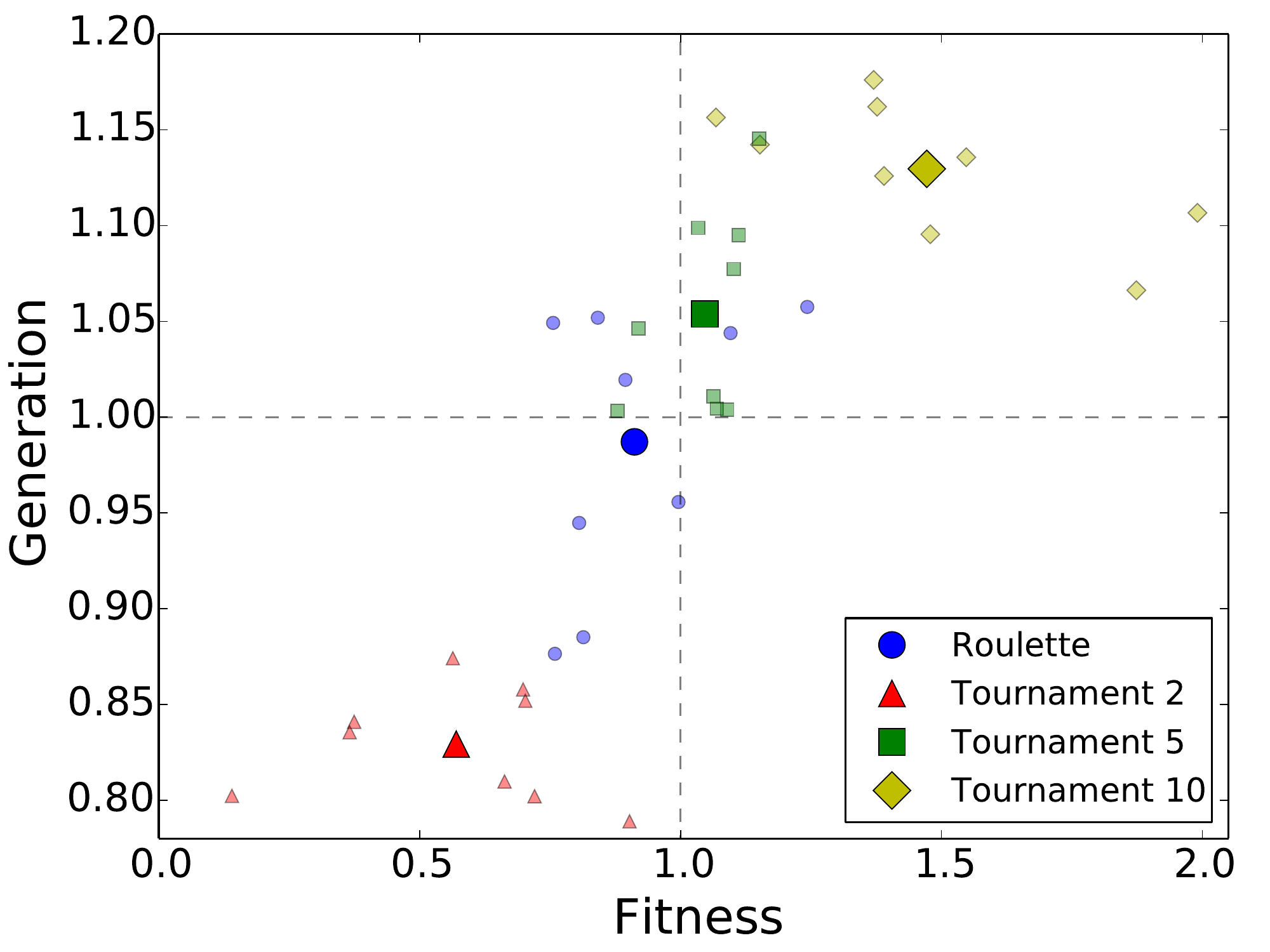}
\captionof{figure}{\it \small A comparison of selection methods for 9 sample datasets. The fitness (measured relative the average performance on that dataset) is plotted against the generation the fitness was achieved (again relative to the average). The large symbols are the centroids of the 9 datasets for each method.}
\label{fig: selection}
\end{Figure}
Fig. \ref{fig: selection} shows the result of a comparison between roulette wheel selection and tournaments of size 2, 5, and 10 (with $P=100$). The horizontal axis shows the best fitness achieved by each method relative to the average of all methods. The vertical axis shows the generation after which the best fitness was achieved relative to the average of all methods. We see that tournaments with only 2 individuals perform worst in terms of fitness. This suggests that the search is less effective which is likely caused by retaining many weak candidates. The best performing selection method examined was tournaments with 10 individuals. This is a result of a more directed search by rewarding the fittest candidates more often. Focusing on the vertical axis we see that smaller tournaments converge to their best fitness earlier than larger tournaments. This is a result of the reduced selection pressure in small tournaments that do not reward fit candidates enough. Large tournaments on the other hand yield a population that was still evolving at the end of this test.

It is natural to think that tournaments that are larger still would yield even better results. This however can lead to problems because if the selection pressure is too high then only the fittest individuals can survive so, the diversity of the population is wiped out and the gene pool stagnates. When this happens we get premature convergence of the GA and since our main interest is in accuracy rather than speed we choose tournaments of 10 individuals for selection.

\subsubsection*{Mutation and Recombination Probability}
Another key parameter in any GA is the mutation rate probability, $p_m$. This controls the ability of a population to generate new solutions. If $p_m$ is too low then the performance of the GA will depend too strongly on the original population because new solutions will be harder to come by. This will often result premature convergence as a result of gene pool stagnation. On the other hand if $p_m$ is too high then we lose a lot of the power of the evolutionary approach. Mutation will destroy good features as well as bad so it will be more difficult for offspring to inherit the good traits of their parents. Both of these extremes can be seen in Fig. \ref{fig: mutation} where we show the dependence of the average fitness on $p_m$ for polynomials with different numbers of terms. In our case, since the size of the search space varies enormously with the number of terms, the optimum $p_m$ will vary slightly depending on the size of the polynomial. In all cases however we see an intermediate $p_m$ in the range $(10^{-2},10^{-1})$ is preferred. We also see that  lower $p_m$ performs better on the larger search spaces of polynomials with more terms. This is expected and is the motivation for using a GA to explore this space rather than simply using a random search. In our final model we set $p_m=0.03$.
\begin{Figure}
\centering
\includegraphics[width=\linewidth]{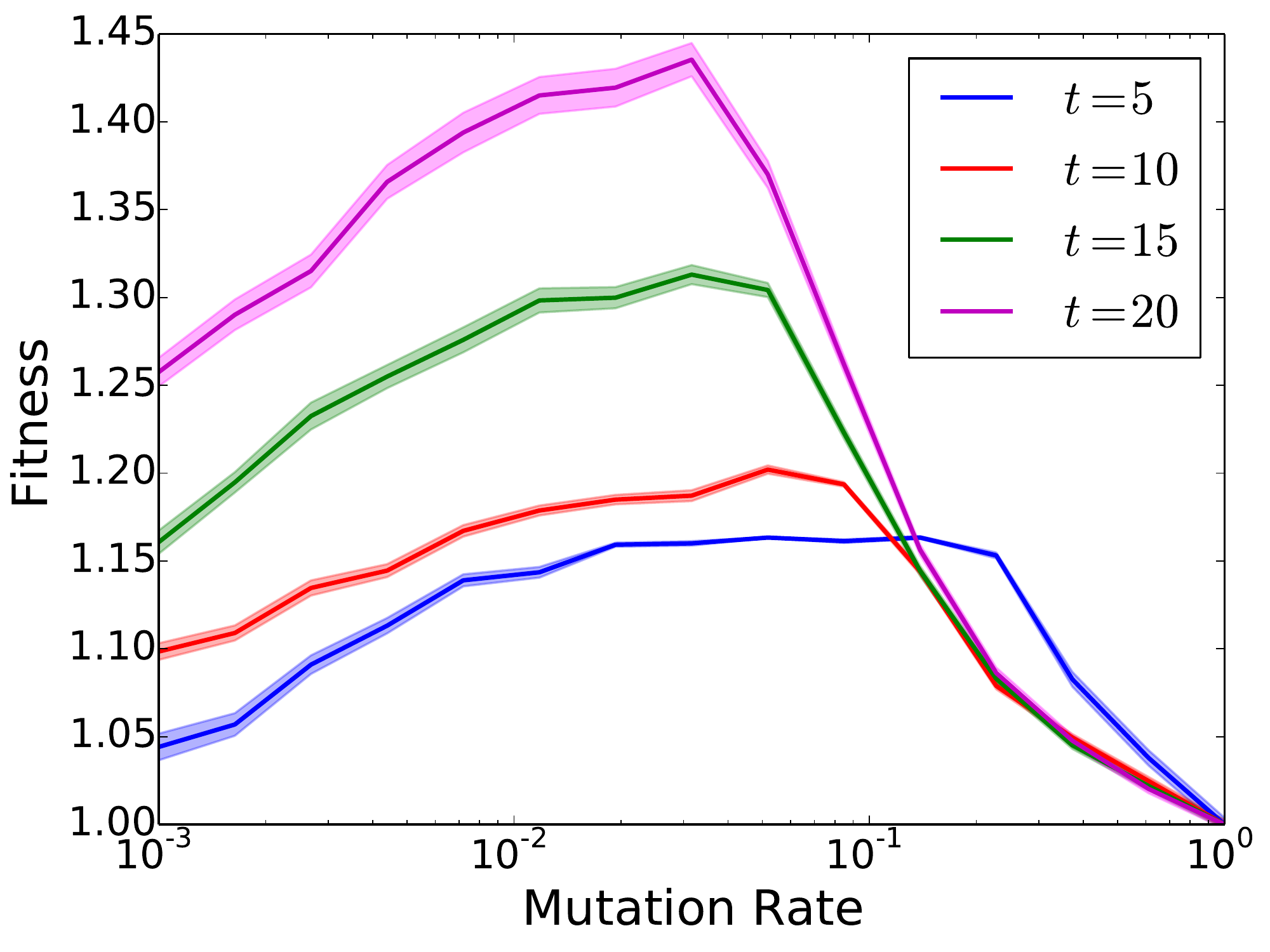}
\captionof{figure}{\it \small Fitness achieved after 100 generations (relative to worst fitness achieve and averaged over 150 runs of the GA) for different values of the mutation probability. $p_m$. The band represents one standard error from the averaging and the results are shown for polynomials with 5, 10, 15, and 20 terms.}
\label{fig: mutation}
\end{Figure}

Yet another tunable parameter of a GA is the recombination probability, $p_r$. This is the probability that two parents, once selected, will breed to produce offspring that proceed to the next generation. If they do not breed a clone of the parent is produced (with the possibility of mutation) to carry forward to the next generation. The result of varying $p_r$ is shown in Fig. \ref{fig: recombination}. We can see that at least a small amount of breeding is important but that there is not much benefit in having very large $p_r$. It is in fact less desirable to have $p_r$ too large because this generates many more distinct polynomials that must be fit to the data which is very time consuming so it is more sensible to chose a lower value of $p_r$ and run the GA longer. This is represented by 'Exploration' in Fig. \ref{fig: recombination} which is the number of polynomials sampled normalised by the number for $p_r=0$. In our final model we set $p_r=0.5$ to balance this trade-off.
\begin{Figure}
\centering
\includegraphics[width=\linewidth]{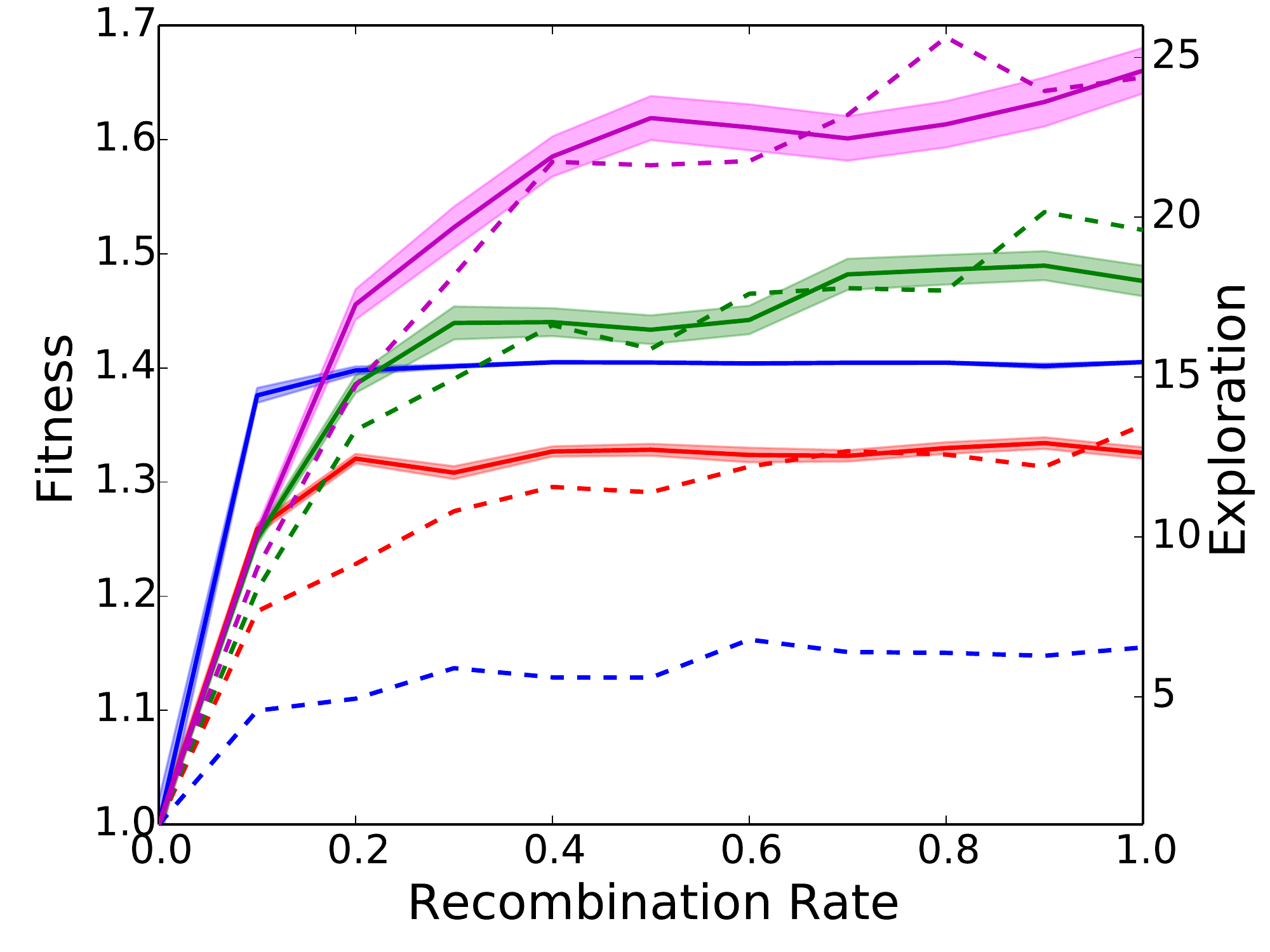}
\captionof{figure}{\it \small Fitness achieved after 100 generations for different values of the recombination probability, $p_m$ (color scheme same as Fig. \ref{fig: mutation}). Also shown is the exploration which measures the number of polynomials investigated relative to case of $p_r=0$.}
\label{fig: recombination}
\end{Figure}
\subsubsection*{Population Size}
Finally, the size of the population needs to be chosen. As one would expect, the result will improve with a larger population but very large populations take more time to evolve and don't make efficient use of the GA's strength which is to evolve a starting population towards better solutions. If the population is too small however then the performance will be limited by the genes of initial population. In this work we chose to use a population of $P=100$ individuals and found that increasing this to larger numbers had only minor effects on the final fitness but dramatically increased run time. For the final result we run for many more generations than in testing and run the GA several times in order to remove dependence on the initial population.

\section{Results: Application to Photometric Redshift Estimation}
In this work we use galaxies with redshifts from the 2SLAQ data set (described in \citep{2SLAQ}) and photometry from the SDSS DR7 \citep{SDSSDR7}. We apply a cut at $i<20$ and select galaxies in the range $0.4 < z <0.7$. This yields $12,306$ distinct objects of which we randomly choose $6,000$ for training. The remaining objects are randomly split into two sets of $3,153$ each for cross validation and testing (we will explain these terms below).
\subsection{Cross Validation}
Cross validation is an imperative step in any machine learning algorithm that helps fight against overfitting. If an algorithm is allowed to fit very complicated functions to the training data then it is common that the learned model will not generalise well to data outside the training set. In order to get a handle on this we use a separate data set for cross validation. The parameters of the model are not trained on the cross validation data so the performance of the model on the this set gives an indication how well the model generalises. The model can then be adjusted in order to yield the best performance on the cross validation set (see Fig. \ref{fig: train_cv_schematic}). Note however that by selecting models based on the cross validation set we have learned from the data so we need yet another data set as a final test that can check the performance of the final model (with optimised parameters).

After cross validation we find the optimal number of terms, $t=20$.  We can also use this technique to choose the degree of the polynomial, $N$. In this instance however the dimension of the search increases dramatically with increasing $N$ so it can be harder to find good solutions and the training error will not necessarily decrease unless the GA is evolved for many more generations. Here we choose $N=5$ because we did not observe any substantial improvement for larger values of $N$ for which the computational cost is much higher. We are therefore searching for the optimal choice of $20$ terms from a possible $252$ terms.
\begin{Figure}
\centering
\includegraphics[width=.9\linewidth]{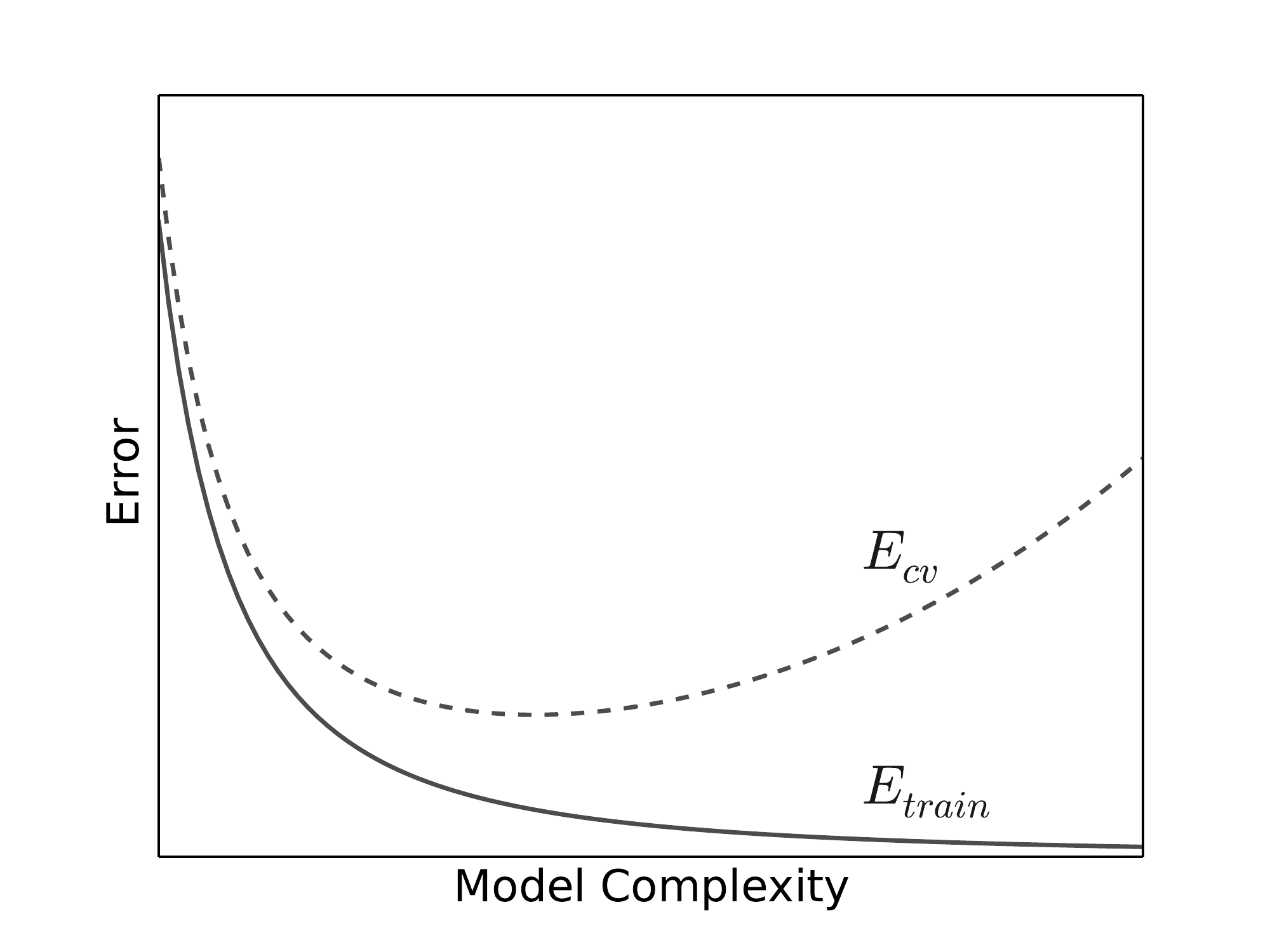}
\captionof{figure}{\it \small Schematic demonstration of cross validation procedure. As the model complexity is increased beyond the optimum it begins to fit noise in the training set and performs worse on cross validation step. The model is chosen in order to minimise validation error.}
\label{fig: train_cv_schematic}
\end{Figure}

\subsection{Test Set Results}
With all of the hyper-parameters now fixed we now have a genetic algorithm specified by the genetic encoding of Eqn. \ref{eqn: genetic encoding} with parameters shown in Table 1. We can now train the model on the training data and test the final trained model on the test set. The resulting best polynomial after multiple runs of the algorithm is shown in the Appendix. The performance of this model is shown visually in Fig. \ref{fig: density plot} where we have plotted logarithmic densities of the predicted redshift of a galaxy, $z_{\text{phot}}$, against the true redshift, $z_{\text{spec}}$, and the diagonal black line represent perfect agreement.

\begin{center}
 \captionof{table}{Final GA parameters}
\begin{tabular}{llr}
\toprule 
\multicolumn{1}{c}{Parameter} &  \multicolumn{1}{c}{\hspace{0.5cm}Value} \\
\midrule
\multicolumn{1}{c}{Selection} & \multicolumn{1}{c}{\hspace{0.5cm}Tournaments of $10$}  \\
\multicolumn{1}{c}{$p_m$}     &  \multicolumn{1}{c}{\hspace{0.5cm}$0.03$}       \\ \multicolumn{1}{c}{$p_r$}      &  \multicolumn{1}{c}{\hspace{0.5cm}$0.5$}       \\ 
\multicolumn{1}{c}{$P$}      &  \multicolumn{1}{c}{\hspace{0.5cm}$100$}       \\ 
\multicolumn{1}{c}{$t$}        &  \multicolumn{1}{c}{\hspace{0.5cm}$20$}                \\ 
\multicolumn{1}{c}{$N$}        &  \multicolumn{1}{c}{\hspace{0.5cm}$5$}                  \\
\bottomrule
 \end{tabular}
\end{center}
\begin{Figure}
\centering
\includegraphics[width=\linewidth]{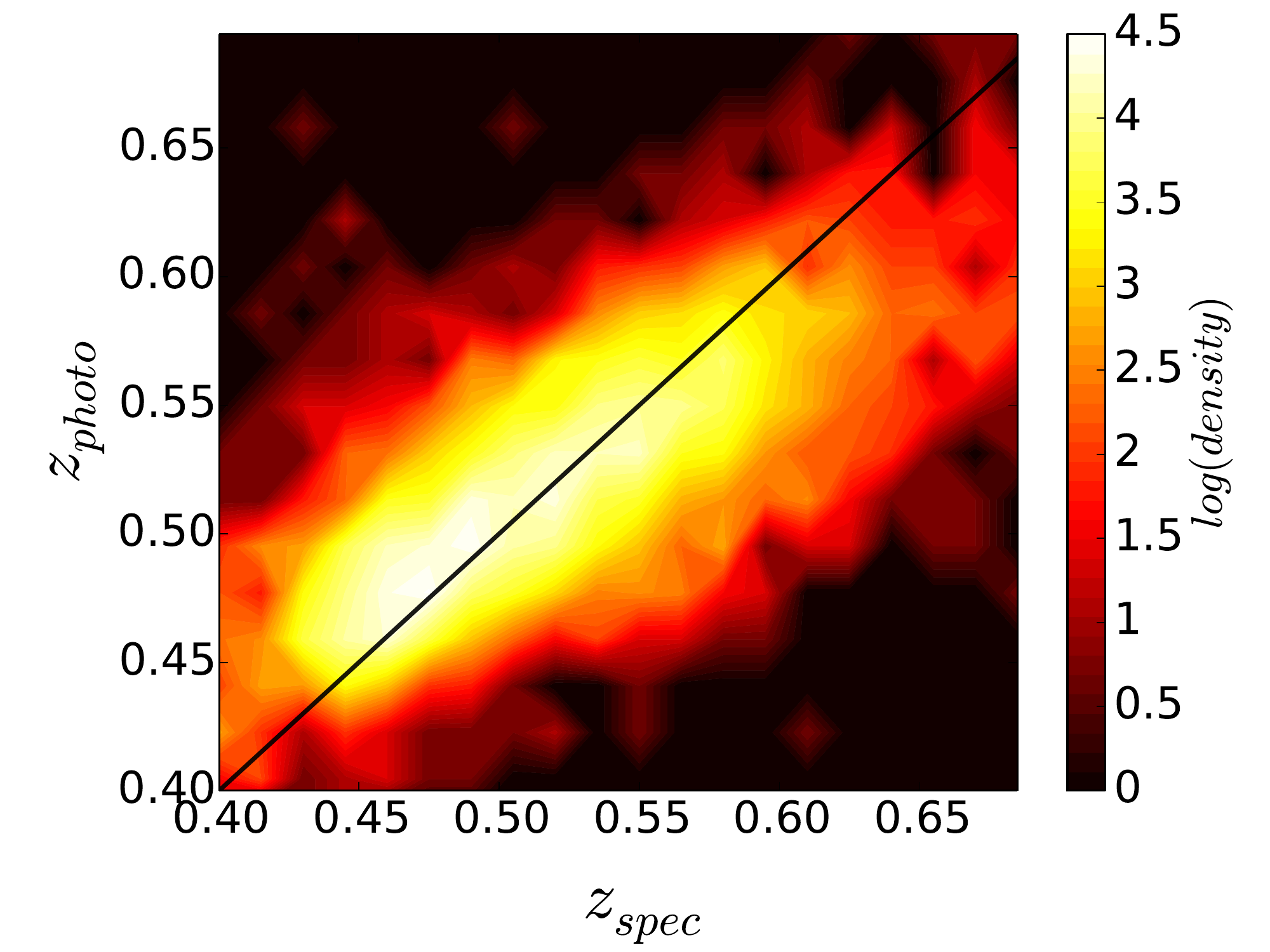}
\captionof{figure}{\it \small Logarithmic density of the true vs. predicted redshifts of the $\sim 3,000$ galaxies in the 2SLAQ LRG test sample}
\label{fig: density plot}
\end{Figure}
There are various metrics to quantify the accuracy of test set performance. For ease of comparison with other approaches to this problem with use the root-mean-squared error,
\begin{equation}
\sigma_{\text{rms}}=\sqrt{\frac{1}{N}\sum_i \left( z_{\text{spec}}^i-z_{\text{phot}}^i \right)^2},
\end{equation}
and we find $\sigma_{\text{rms}}=0.0414$ for the $3,153$ galaxies in the test set. This result compares favourably with those studied in \citep{Abdalla0812} where the results lie in the range $0.057 <\sigma_{\text{rms}} <0.097$ with ANN$z$ performing the best (Note that these test were performed with data in the slightly extended redshift range $0.3 <z <0.8$, in which case the method present here yields\footnote{The decreased performance is largely due to lack on training data with $z<0.4$ and $z>0.7$} a $\sigma_{\text{rms}}=0.052$). It must be noted that these tests were carried out using SDSS DR6 photometry and there was a significant improvement in photometric data from DR6 to DR7 \citep{SDSSDR7}. For the case of ANN$z$ however it was shown in \citep{Thomas2011} that using DR7 photometry did not change $\sigma_{\text{rms}}$ indicating that the difference may not be solely attributable to photometry.

Another interesting metric to consider is the catastrophic outlier rate, $\mathcal{O}_{0.1}$, defined a the percentage of test set galaxies with $|z_{\text{spec}}-z_{\text{phot}}|>0.1$. For the 2SLAQ sample we find $\mathcal{O}_{0.1}=2.70\% $

The training data is not distributed uniformly in redshift with a large portion lying in the range $0.45 < z_{\text{spec}} < 0.55$. We would therefore expect that the algorithm attaches more importance to correctly predicting the redshift in this range. The expectation is borne out in the resulting $z_{\text{spec}}$ dependence of $\sigma_{\text{rms}}$ shown in Fig. \ref{fig: z breakdown of error} (a) where we see $\min(\sigma_{\text{rms}}) \simeq 0.02$ lies at $z_{\text{spec}}\simeq 0.53$.
\begin{Figure}
\centering
\includegraphics[width=\linewidth]{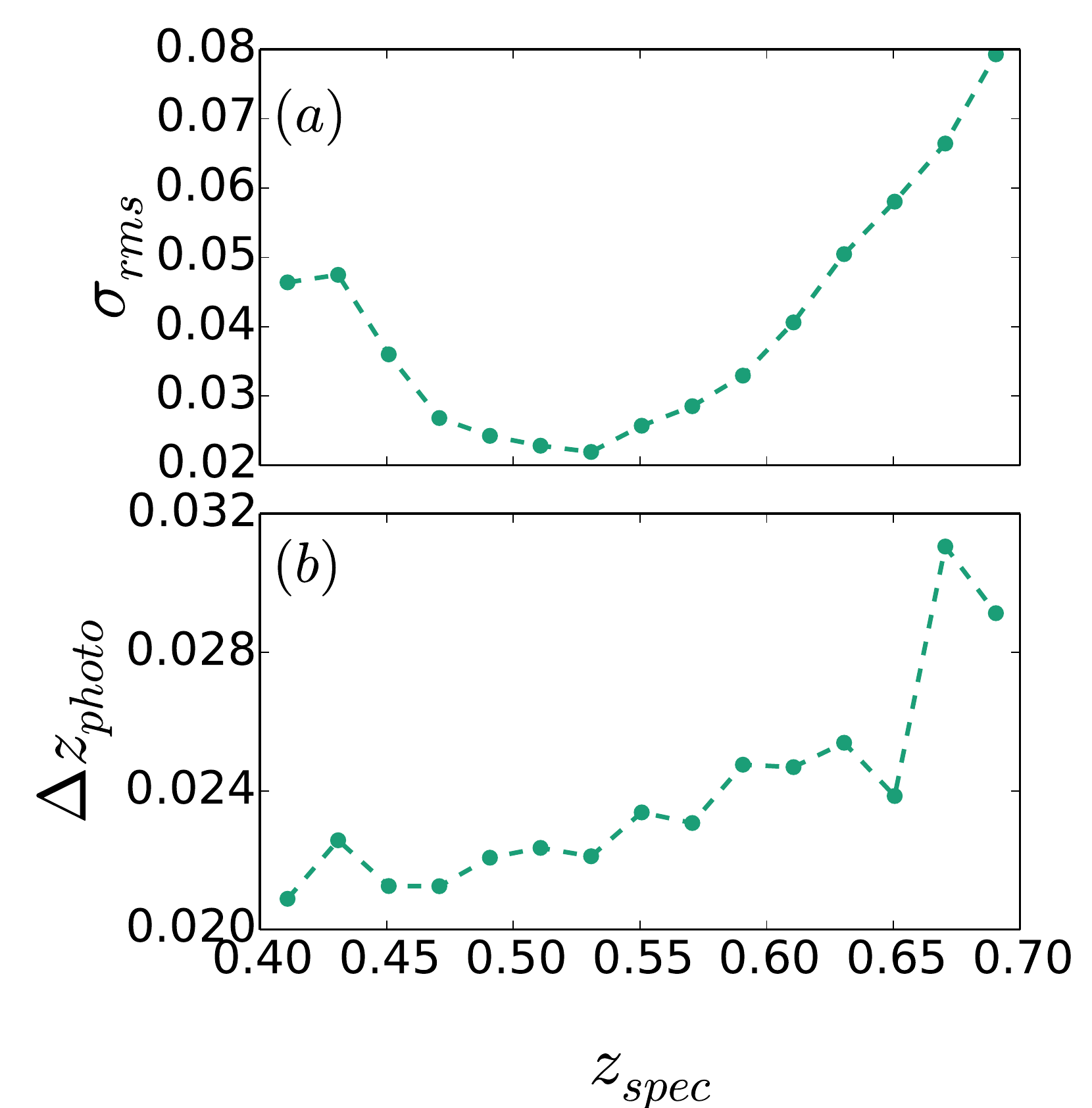}
\captionof{figure}{\it \small (a) Redshift dependence of rms error. The error is minimised in the redshift range where most of the training data lies. (b) The prediction error estimate coming from experimental errors on photometric data. The error grows for larger redshifts and this is reflected in the larger rms error for galaxies in this region.}
\label{fig: z breakdown of error}
\end{Figure}
The effect of experimental errors of the photometry can also be taken into account at the results stage. Given a polynomial model we can easily propagate through the photometric errors to produce an error on the predicted photometric redshift,
\begin{equation}
\left(\Delta z_{\text{phot}}\right)^2=\sum_i \left( \frac{\partial z_{\text{phot}}(m_i)}{\partial m_i}\right)^2 \left(\Delta m_i \right)^2,
\end{equation}
where $m_i=\{u,g,r,i,z\}$. The redshift dependence of this error can be seen in Fig. \ref{fig: z breakdown of error} (b). We see that at larger redshifts the photometric data worsens and results in a larger error on the predict photometric redshift. To illustrate the effect of photometric errors we have plotted a representative subsample of $300$ galaxies and their errorbars in Fig. \ref{fig: errorbars}. We see in Fig. \ref{fig: density plot} and Fig. \ref{fig: errorbars} that the model tends to overestimate the redshift and low $z_{\text{spec}}$ and underestimates it at high $z_{\text{spec}}$.
It would be nice to use the error information for training so that galaxies with worse photometric data are weighted less in the fit. For example one could minimise the $\chi^2$,
\begin{equation}
\chi^2 =\sum_{i} \frac{( z^i_{\text{phot}}-z^i_{\text{spec}})^2}{(\Delta z^i_{\text{phot}})^2}=\sum_{i} \frac{( \delta z^i)^2}{(\Delta z^i_{\text{phot}})^2}.
\end{equation}
This, however, does not work because there will be a degeneracy between decreasing $\delta z^i$ and increasing $\Delta z^i_{\text{phot}}$. This means that we might end up with models that have poor predicting power. Unfortunately therefore we must neglect this information and hope that cross validation will prevent overfitting to galaxies with large photometric errorbars.
\begin{Figure}
\centering
\includegraphics[width=\linewidth]{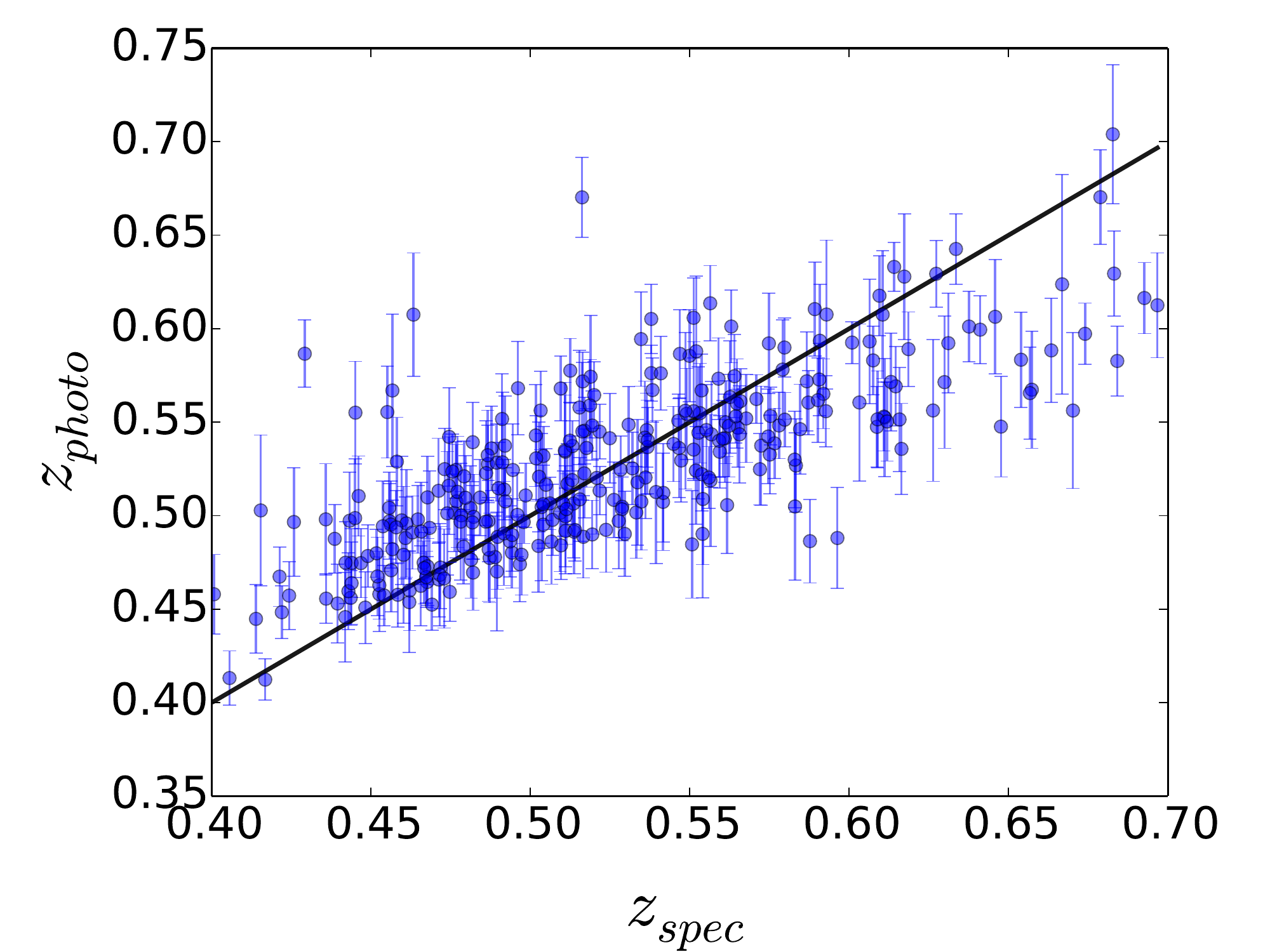}
\captionof{figure}{\it \small Photometric redshift prediction and errorbars for a representative subsample of $300$ galaxies. The errorbars are due to errors in the photometric data and so depend on the particular model chosen for $z_{\text{phot}}$.}
\label{fig: errorbars}
\end{Figure}
In order to estimate the dependence of this result on the choice of training, cross-validation, and test sets we could repeat the entire procedure for many other random choices. Since we used the data to fix hyper-parameters in principle these should be fit again each time. This would be a time consuming process so we instead take the hyper-parameters as fixed (see Table 1) and perform the test on the final GA. We therefore do not need a cross-validation step so we randomly select $6,000$ galaxies for training and $3,153$ galaxies for testing and run the GA 5 for each selection. The distribution of $\sigma_{\text{rms}}$ is shown in Fig. \ref{fig: error dist}. The distribution  is narrow suggesting that the dependence on the choice of data sets is minimal\footnote{Note that some of the spread could be due to the GA not reaching the optimal solution for a particular dataset but we try to mitigate against this by running for 500 generations 5 separate times.}. We can set a tentative error estimate on $\sigma_{\text{rms}}$ using the standard deviation of this distribution and find $\sigma_{\text{rms}}=0.0408 \pm 0.0006$.
\begin{Figure}
\centering
\includegraphics[width=\linewidth]{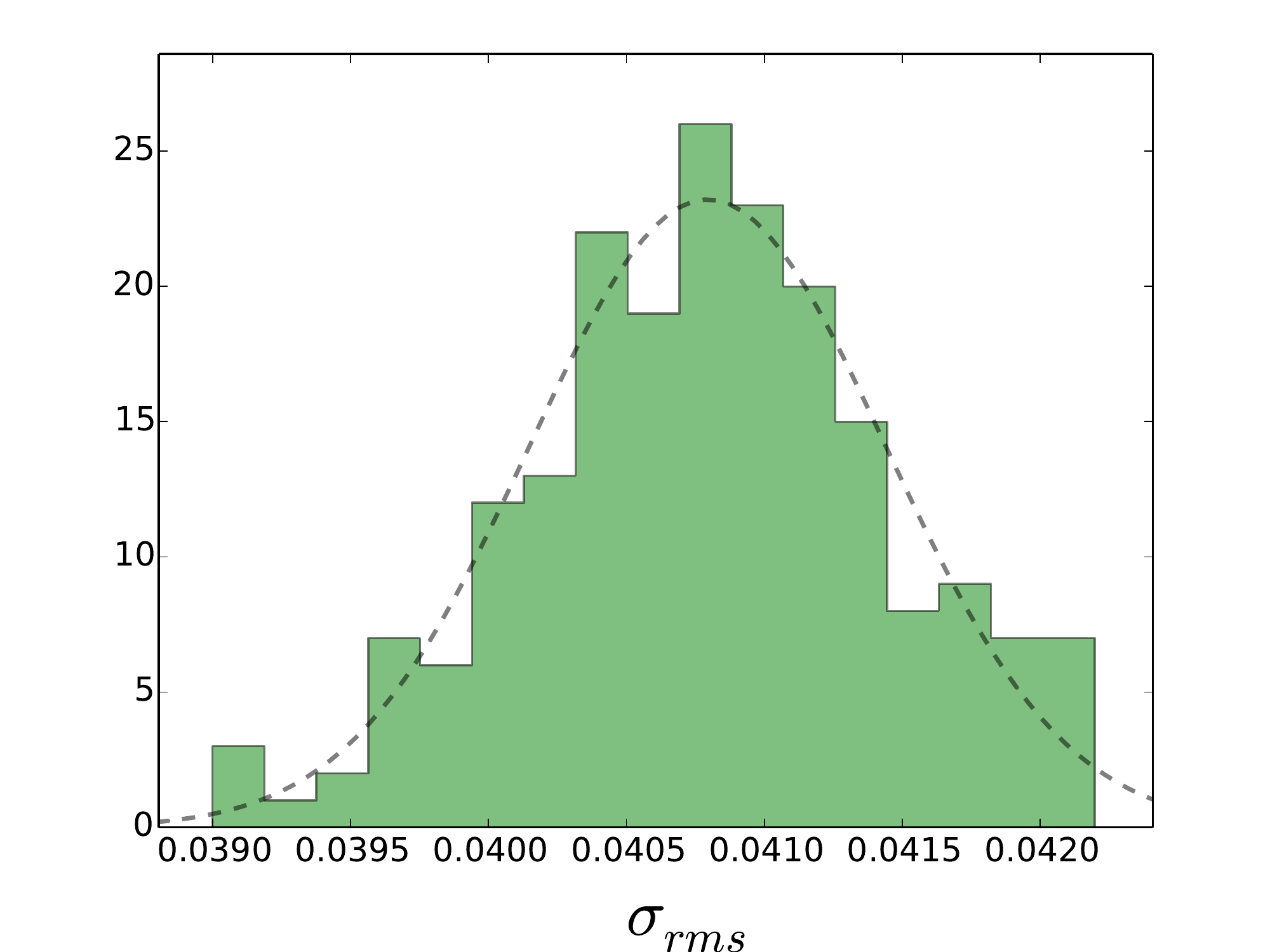}
\captionof{figure}{\it \small Distribution of photometric error obtained by repeatedly selecting training and test sets at random. Results shown for 200 repetitions and the best fit gaussian distribution is included from comparison.}
\label{fig: error dist}
\end{Figure}
\subsection{Other Data Sets}
In order to further test the performance of the method we have applied the GA (with all hyper-parameters fixed) to other data sets with different redshift ranges. Firstly, we use SDSS DR11 data to get a large sample of almost $700,000$ galaxies\footnote{The query used for the SDSS CasJobs server is that described in \citep{Kind2014,GLM}} in the range $0< z_{spec} <1$ (Note that we do not perform any magnitude cuts on this sample). We use $50,000$ galaxies to train the GA and $\sim 630,000$ to test it with the remainder used for cross validation. The result is shown in Fig. \ref{fig: DR11} and corresponds to a $\sigma_{\text{rms}}=0.0336$ with a catastrophic outlier rate of $ \mathcal{O}_{0.1}=1.02\% $. These results again compare well with other methods on the same data set \citep{Kind2014,GLM} (different metrics are used in these works so we do not make a direct comparison here). 

We also applied the method to the PHAT0 data set from \citep{Hildebrandt1008} which contains $\sim 170,000$ galaxies. This is synthetic data based on simulations to which some noise is added. The data was generated using LePhare \citep{LePhare1} using 11 different filter magnitudes. We therefore train a GA with 11 variables to reproduce the redshift. We use $\sim 34,000$ galaxies to train the GA and $\sim 120,000$ to test it with the remainder used for cross validation. The hand added noise in this data set is much lower than that of real data so it is possible to include many more terms in the polynomial before overfitting. The result of a fit with 200 terms in shown in Fig. \ref{fig: Phat0}. This corresponds to a $\sigma_{\text{rms}}=0.0188$ with a catastrophic outlier rate of $ \mathcal{O}_{0.1}=0.096\% $. 
\begin{Figure}
\centering
\includegraphics[width=\linewidth]{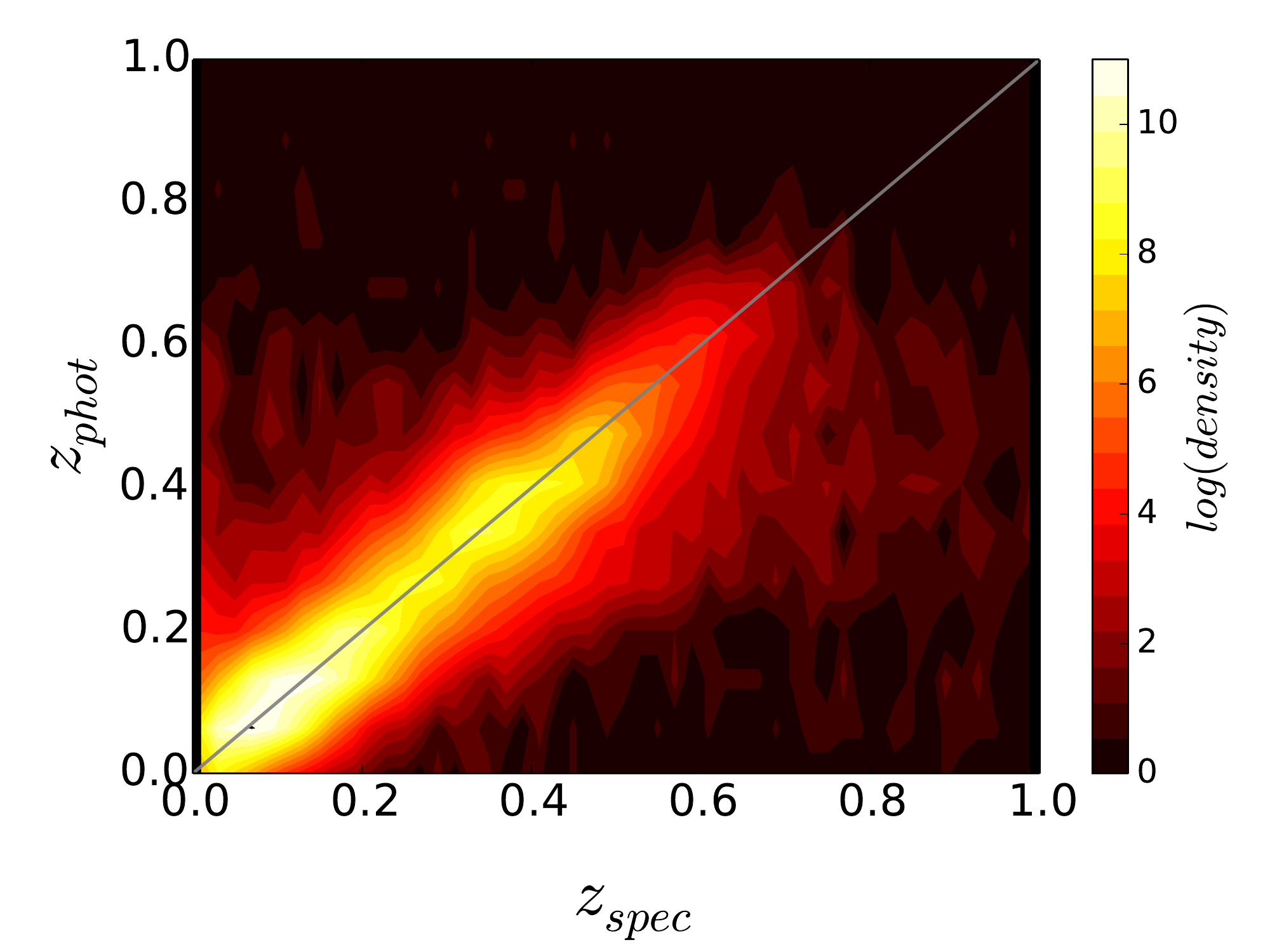}
\captionof{figure}{\it \small Logarithmic density of the true vs. predicted redshifts of the $\sim 630,000$ galaxies in the SDSS DR11 test sample}
\label{fig: DR11}
\end{Figure}
\begin{Figure}
\centering
\includegraphics[width=\linewidth]{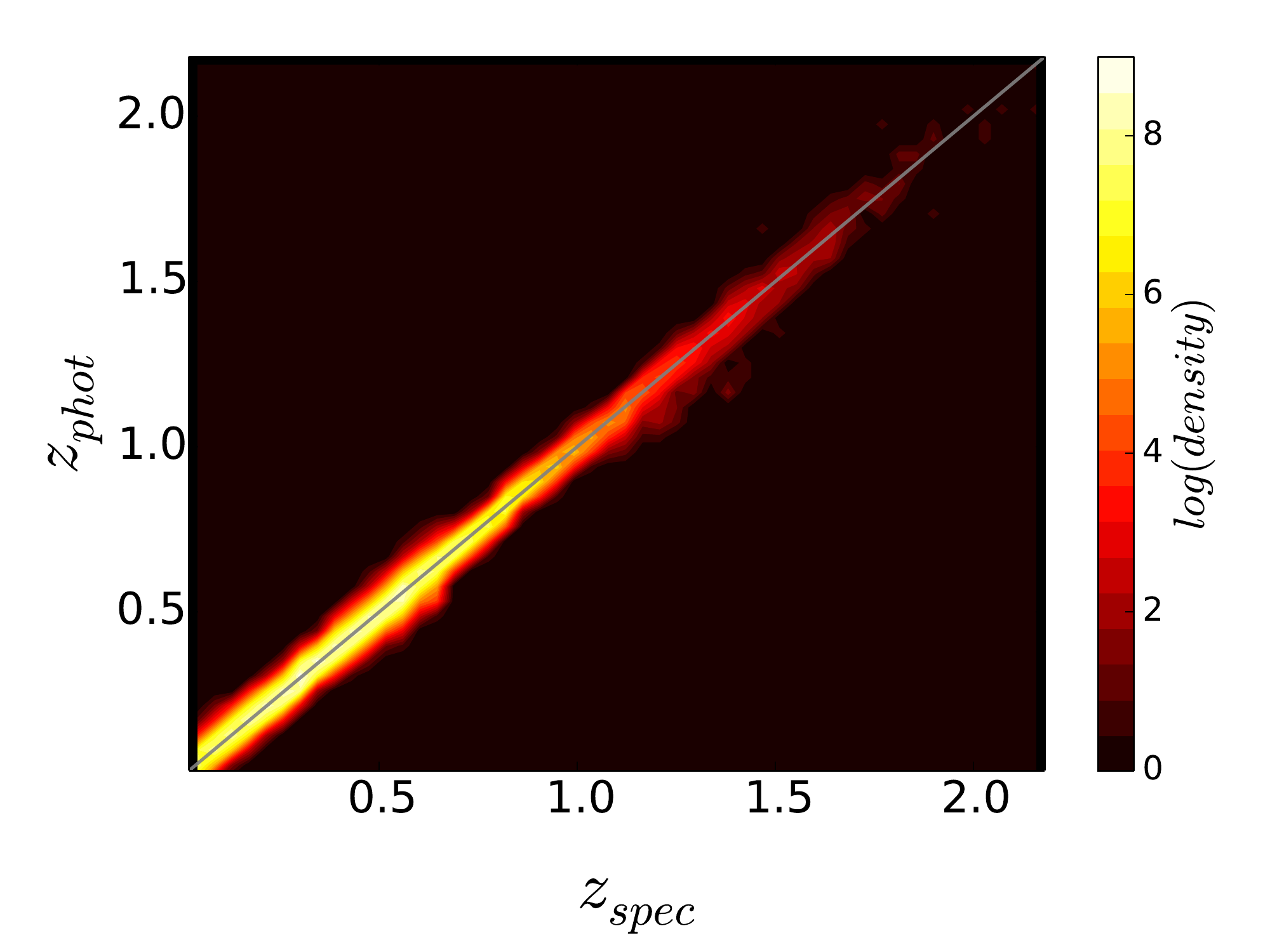}
\captionof{figure}{\it \small Logarithmic density of the true vs. predicted redshifts of the $\sim 120,000$ galaxies in the PHAT0 test sample}
\label{fig: Phat0}
\end{Figure}

In both cases the GA has performed well demonstrating that the method generalises well to new data sets with larger redshift ranges.

\subsection{Code Performance}
In this section we briefly mention the performance of the code and how it scales with training set size. It is important that a photo-z code can train on a set of $\sim 10^5$ galaxies in a reasonable amount of time (although training on the full set is only necessary once for real world applications). In Fig. \ref{fig: performance} we show the time taken to train for 500 generations on various data set sizes from DR11 for different numbers of terms. The solid lines represent linear scaling with training set size\footnote{Tests were carried out on a Intel Core i7-2600 3.40 GHz system with 11.7 GB RAM running Scientific Linux}. We see that the performance scales linearly with training set size and takes approximately half and hour for a typical training set size of $\sim 30,000$ galaxies with 20 terms. If even faster performance is required then it would be possible to make use of the readily parallelisable nature of GAs. Although it has not been implemented in this version of the code one could in principle speed up the performance by a factor $\sim 100$ (the population size) if the fitness evaluation of each individual was computed simultaneously on different cores.

\begin{Figure}
\centering
\includegraphics[width=\linewidth]{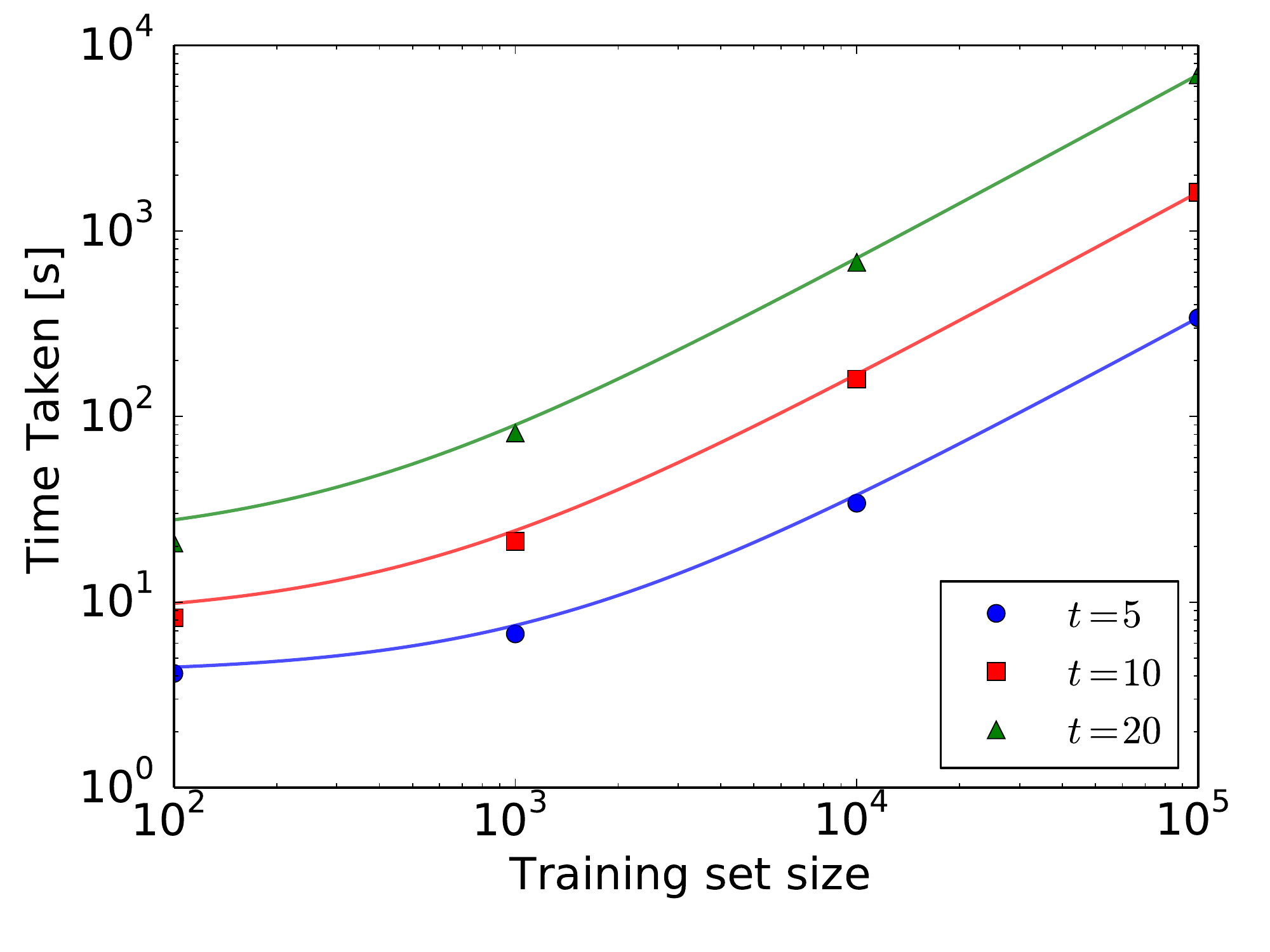}

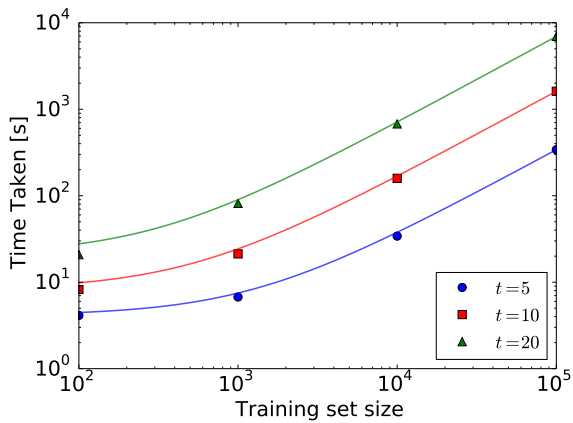
\captionof{figure}{\it \small Performance analysis for runs of GA with 500 generations on the DR11 training set with different numbers of training examples and different numbers of terms. The solid lines shows linear scaling with training set size.}
\label{fig: performance}
\end{Figure}

\section{Conclusions}
In this paper we have demonstrated the use of a genetic algorithm to explore the space of polynomial models for photometric redshift estimation. We have found $\sigma_{\text{rms}}=0.0408\pm 0.0006$ in the redshift range $0.4 < z < 0.7$ on a test set of more than $3,000$ 2SLAQ galaxies which was not used at any stage of model training. This result compares favourably with current methods in the literature. We have also shown that the method performs very well on other data sets (DR11 and PHAT0) with varied redshift ranges. It will be interesting to see a more in depth comparison with other codes with different figures of merit and data sets, for example simulated data for the Euclid \citep{Euclid} and LSST\citep{LSST} surveys, which we will leave for future work.

The success of the model is likely due to the small number of degrees of freedom in the final model (so overfitting can be avoided) while still maintaining the flexibility to adapt well to the training set using the GA. The resulting best fit polynomial has been presented and can be easily used to generate photometric redshift catalogues without the need to run any additional code.
\section*{Acknowledgements}
We are grateful to comments and suggestions from Filipe Abdalla, Philip Grothaus and Jon Loveday.  The work of RH is funded by a grant from the KCL graduate school while MF receives funding from the Science and Technology Facilities Council.

\bibliographystyle{mn2e}
\bibliography{GAz.bib}
\end{multicols}
\section*{Appendix}
The best polynomial discovered by the GA is presented below,
\begin{align*}
z_{\text{phot}}(u,g,r,i,z)= &-40.383 + \frac{1}{m} \left(0.832 u -47.44 g+357.4 r-91.44 z \right)\\
&+\frac{1}{m^2} \left(49.88 g r -960.8 r^2 \right) \\
&+\frac{1}{m^3} \left(-18.396 g^3+1299.2 g^2 i -1512.0 g i^2+82.3 r^2 i+325.44 r^2 z+551.2 i^3\right)\\
&+\frac{1}{m^4}\left( -4.293 u i z^2 -247.52 r z^3+115.12 z^4\right)\\
& +\frac{1}{m^5}\left(-0.32 u^3 r z +1.754 u^2 g z^2 -5.859 g^3 r z+16.179 i z^4 \right)
\end{align*}
where $m=20$ is a reference magnitude.
\end{document}